\documentclass[aps,prd,preprint,floatfix,superscriptaddress,nofootinbib,longbibliography,a4paper]{revtex4-1}
\pdfoutput=1
\usepackage{color}
\usepackage{graphicx}
\usepackage{textcomp}
\usepackage{subfigure}
\usepackage{feynmp}
\usepackage{amsmath,amssymb,array}
\usepackage{enumerate}
\usepackage{hhline}
\newcommand{\mathsym}[1]{{}} 

\def\gsim{\:\raisebox{-1.1ex}{$\stackrel{\textstyle>}{\sim}$}\:}

\newcommand{\beqa}{\begin{eqnarray}}
\newcommand{\eeqa}{\end{eqnarray}}
\newcommand{\be}{\begin{equation}}
\newcommand{\ee}{\end{equation}}
\newcommand{\ba}{\begin{array}} 
\newcommand{\ea}{\end{array}}

\begin{document} 
\vspace*{0.5cm}
\title{Leptogenesis and fermion mass fit in a renormalizable $SO(10)$ model}
\bigskip
\author{V. Suryanarayana Mummidi}
\email{venkata@nitt.edu}
\affiliation{Department of Physics, National Institute of Technology, Tiruchirappalli-620 015, India.}
\author{Ketan M. Patel}
\email{kmpatel@prl.res.in}
\affiliation{Theoretical Physics Division, Physical Research Laboratory, Navarangpura, Ahmedabad-380 009, India. \vspace*{0.5cm}}

\begin{abstract}
A non-supersymmetric renormalizable $SO(10)$ model is investigated for its viability in explaining the observed fermion masses and mixing parameters along with the baryon asymmetry produced via thermal leptogenesis. The Yukawa sector of the model consists of complex $10_H$ and $\overline{126}_H$ scalars with a Peccei-Quinn like symmetry and it leads to strong correlations among the Yukawa couplings of all the standard model fermions including the couplings and masses of the right-handed (RH) neutrinos. The latter implies the necessity to include the second lightest RH neutrino and flavor effects for the precision computation of leptogenesis. We use the most general density matrix equations to calculate the temperature evolution of flavoured leptonic asymmetry. A simplified analytical solution of these equations, applicable to the RH neutrino spectrum predicted in the model, is also obtained which allows one to fit the observed baryon to photon ratio along with the other fermion mass observables in a numerically efficient way. The analytical and numerical solutions are found to be in agreement within a factor of ${\cal O}(1)$. We find that the successful leptogenesis in this model does not prefer any particular value for leptonic Dirac and Majorana CP phases and the entire range of values of these observables is found to be consistent. The model specifically predicts (a) the lightest neutrino mass $m_{\nu_1}$ between 2-8 meV, (b) the effective mass of neutrinoless double beta decay $m_{\beta \beta}$ between 4-10 meV, and (c) a particular correlation between the Dirac and one of the Majorana CP phases.
\end{abstract}

\maketitle
\section{Introduction}
\label{sec:intro}
Augmentation of the Standard Model (SM) with gauge singlet fermions - the so-called Right-Handed (RH) neutrinos - provides a natural explanation for the observed small masses of weakly interacting neutrinos through the seesaw mechanism \cite{Minkowski:1977sc,Yanagida:1979as,Mohapatra:1979ia,Schechter:1980gr}. Generating matter-antimatter asymmetry through leptogenesis \cite{Fukugita:1986hr} (see also \cite{Bodeker:2020ghk,DiBari:2021fhs,Xing:2020ald} for the recent reviews) is a direct cosmological application of the RH neutrinos. The existence of these new fermions is naturally predicted by a class of Grand Unified Theories (GUT) based on $SO(10)$ gauge symmetry \cite{Fritzsch:1974nn,GellMann:1980vs}. Along with the unification of strong and electroweak interactions, it also provides a complete unification of the SM  quarks and leptons of a given generation in a single irreducible spinorial representation of the underlying gauge group which also includes an RH neutrino. Therefore, the masses and couplings of the RH neutrinos are related with those of the SM fermions. The exact nature of relations depend on the specific scalar sector one considers for a given version of $SO(10)$ GUT.

The Yukawa sector of renormalizable non-supersymmetric $SO(10)$ GUTs has been extensively investigated for its viability in reproducing the low-energy data of fermion masses and mixing parameters \cite{Babu:1992ia,Bajc:2005zf,Joshipura:2011nn,Altarelli:2013aqa,Dueck:2013gca,Meloni:2014rga,Meloni:2016rnt,Babu:2016bmy,Ohlsson:2018qpt,Boucenna:2018wjc,Ohlsson:2019sja}. All the fermions residing in three generations of $16$-plet can interact with Lorentz scalar $10_H$, $120_H$ and $\overline{126}_H$ dimensional representations of $SO(10)$. The first two of these can be chosen as real or complex. Minimum two scalar fields are necessary to reproduce a realistic spectrum of quarks and leptons \cite{Babu:1992ia}. Additional restrictions like a Peccie-Quinn symmetry \cite{Bajc:2005zf,Joshipura:2011nn} and/or condition of spontaneous CP violation \cite{Joshipura:2011nn} are also often used to reduce the number of Yukawa couplings in order  to construct more predictive models. These restrictions allow one to write down the charged fermions and neutrino mass matrices in terms of very few fundamental Yukawa couplings and Vacuum Expectation Values (VEV) of scalars and give rise to correlations between the quark and lepton spectrum. These correlations are then used to first check the viability of the underlying model, and if found to be consistent with the known data, to derive predictions for observables that are not yet determined in the experiments. These predictions can be used to establish further the validity of the underlying model or otherwise. For example, based on this procedure a specific range for the reactor neutrino mixing angle was predicted in a model with complex $10_H$ and $\overline{126}_H$ scalars in \cite{Joshipura:2011nn} which was found in excellent agreement with its value independently measured by Daya Bay \cite{DayaBay:2012fng} and RENO \cite{RENO:2012mkc} experiments a year later.

Since RH neutrinos are naturally accommodated with the other SM quarks and leptons in $SO(10)$ GUTs, possibility of leptogenesis in these frameworks has also been explored in several works \cite{Buchmuller:1996pa,Nezri:2000pb,Buccella:2001tq,Branco:2002kt,Akhmedov:2003dg,DiBari:2008mp,DiBari:2010ux,Buccella:2012kc,DiBari:2014eya,Fong:2014gea,DiBari:2015oca,DiBari:2017uka,DiBari:2020plh}. In these studies, $SO(10)$-inspired relationship between the Dirac neutrino and up-type quark mass matrices is assumed. This allows complete or partial determination of the RH neutrino mass spectrum from the light neutrino masses and mixing leading to a predictive setup to study the thermal leptogenesis. As one of the important results, it has been pointed out that in most of the cases the lepton asymmetry is dominantly generated by the decays of the second lightest RH neutrino \cite{Vives:2005ra,DiBari:2005st,Abada:2006fw,Abada:2006ea,Nardi:2006fx,Engelhard:2006yg}. This is referred as $N_2$ dominated leptogenesis and the dependence of the lepton flavors on CP asymmetries and washout processes play crucial role in this scenario. Subsequently, the conditions for successful leptogenesis in $SO(10)$-inspired models have been extensively studied both analytically and numerically and low energy predictions for the light neutrino observables have been derived in \cite{DiBari:2013qja,DiBari:2017uka,Chianese:2018rnq,DiBari:2020plh}.

While leptogenesis has been studied using certain $SO(10)$-inspired conditions, its comprehensive analysis in concrete and realistic $SO(10)$ models has not been carried out in the necessary details. As it will be described later in this paper, the constrained Yukawa sector of realistic and predictive $SO(10)$ models do not relate only the Dirac neutrino Yukawa couplings with those of the up-type quarks but also relate the RH neutrino mass matrix with the mass matrices of other SM fermions. Therefore, the robust evaluation of baryon asymmetry needs to be carried out in the region of parameter space allowed by the viable fermion mass spectrum. To the best of our knowledge, most of the fermion mass fits performed earlier in realistic $SO(10)$ models do not include leptogenesis except \cite{Altarelli:2013aqa}. In Ref. \cite{Altarelli:2013aqa}, the final lepton asymmetry is evaluated using an approximate analytical solution of Boltzmann equations governing the evolution of $B-L$ asymmetry produced mainly from the decay of the lightest RH neutrino. It is then fitted with the other fermion masses and mixing angles. As we show  in this paper, the RH neutrino mass spectrum and Dirac neutrino Yukawa couplings allowed by realistic fermion mass spectrum are such that the lepton asymmetry is dominantly generated by the decays of the second lightest RH neutrino. It also requires careful incorporation of lepton flavor effects beyond the simplified approach adopted in \cite{Altarelli:2013aqa}. Therefore, investigation of viable leptogenesis requires more precise computation of evolution of flavor specific lepton asymmetries produced from the decays of all the three RH neutrinos and subsequent washouts effects. This is the main aim of the present work.

The recent advances in computing kinetic evolutions of flavor specific lepton asymmetries using Density Matrix Equations (DME) instead of the usual Boltzmann Equations (BE) allow more accurate evaluation of the final lepton asymmetry for arbitrary RH neutrino mass spectrum \cite{Blanchet:2011xq}. We use these equations to first derive an approximate analytical solution applicable for the RH neutrino mass spectrum and Yukawa couplings predicted by an $SO(10)$ model with $10_H$ and $\overline{126}_H$ Higgs. This solution has a simple form and it is quite useful for fitting the baryon to photon ratio $\eta_B$ to its observed value along with all the fermion masses and mixing parameters in a numerically very efficient way. Once the values of fundamental Yukawa couplings and VEVs are determined in this way, we also compute $\eta_B$ by numerically solving the full DME. We look for several solutions which are statistically allowed and derive comprehensive predictions for many observables in the lepton sector.

The paper is organized as follows. In the next section, we discuss the Yukawa sector of an $SO(10)$ model with two scalar representations and derive the correlations between the quark and lepton sector including the mass spectrum of RH neutrinos. The detailed treatment of leptogenesis is presented in section \ref{sec:leptogenesis} where we discuss the general DME formalism and its simplification for the obtained mass spectrum of the RH neutrinos. Numerical analysis, results and predictions of the model are discussed in section \ref{sec:numerical}. Finally, the study is summarized in section \ref{sec:summary} and some relevant supplementary material is given in two sections in Appendix.

\section{Yukawa sector of minimal $SO(10)$ and Right-Handed neutrino mass spectrum}
\label{sec:YukawaSector}
In the renormalizable versions of $SO(10)$ models, the masses of all the fermions residing in three generations of $16_F$ can arise from their Yukawa interactions with three scalar representations: $10_H$, $120_H$ and $\overline{126}_H$. For realistic fermion masses and mixing angles, Yukawa interactions with at least two of these scalar multiplets are necessary \cite{Babu:1992ia}. For the present study, we choose a $10_H$ and $\overline{126}_H$ as minimal\footnote{Here, the minimality is defined in the sense of number of Higgs representations used in the Yukawa sector. We have chosen a complex $10_H$. Alternatively, a case for real $10_H$ and $120_H$ along with $\overline{126}_H$ without any additional symmetry has also been put forward as minimal Yukawa sector in \cite{Babu:2016bmy}.} Yukawa sector which is known to provide realistic fermion masses \cite{Joshipura:2011nn,Dueck:2013gca}. The $\overline{126}_H$ can break $U(1)_{B-L}$ subgroup of $SO(10)$ and generate Majorana masses for the SM singlet and the active neutrinos. The most general Yukawa interactions with a complex $10_H$ and $\overline{126}_H$ can be written as
\be \label{LY_GUT}
-{\cal L}_Y = 16_F^T\,\left(Y_{10}\, 10_H\,+\,\tilde{Y}_{10}\,10_H^*\,+\, Y_{126}\,\overline{126}_H\right) 16_F + {\rm h.c.} \,.\ee
Here, $Y_{10}$, $\tilde{Y}_{10}$ and $Y_{126}$ are symmetric matrices in generation space.

Different SM fields residing in the above representation can be identified as
\beqa \label{decomposition}
16_F &=& Q_L (3,2,1/6) + u^c_L (\overline{3},1,-2/3) + d^c_L (\overline{3},1,1/3)\nonumber \\
&+& L_L (1,2,-1/2) + e^c_L (1,1,1) + \nu^c_L (1,1,0)\,,\, \nonumber\\
10_H &=&  h_{10,u} (1,2,-1/2) + h_{10,d} (1,2,1/2) + ... \,,\nonumber \\
\overline{126}_H &=& h_{126,u} (1,2,-1/2) + h_{126,d} (1,2,1/2) + S(1,1,0) + ... \,,
\eeqa
where, the numbers in parentheses denote the $SU(3)_C$, $SU(2)_L$ and $U(1)_Y$ charges, respectively. As it can be seen, each $10_H$ and $\overline{126}_H$ contains a pair of Higgs doublet with opposite hypercharge. Additional doublets may also arise from the fields like $210_H$ which do not contribute to the Yukawa interactions but may be needed to break the unified gauge group \cite{Bertolini:2009es,Bertolini:2009qj}. Once $SO(10)$ is broken into the SM, the different doublets of the same hypercharge get mixed with each other and it is typically assumed that only a pair of them remains light and it breaks electroweak symmetry at the scales much below the GUT scale. We denote this light pair by $h_{u,d}$ and  parametrize the different doublets of $10_H$ and $\overline{126}_H$ as
\beqa \label{huhd}
h_{10,d} &=& \alpha_1 h_d\,,~h_{126,d} = \alpha_2 h_d\,, \nonumber \\
h_{10,u} &=& \beta_1 h_u\,,~h_{126,u} = \beta_2 h_u\,. \eeqa
The parameters $\alpha_i$ and $\beta_i$ can be computed in terms of the parameters of the Lagrangian once the full scalar spectrum of the theory is specified. In general, $|\alpha_1|^2 + |\alpha_2|^2 \le 1$ and $|\beta_1|^2 + |\beta_2|^2 \le 1$. The equality holds when the Higgs doublets in the theory arise from only $10_H$ and $\overline{126}_H$.

After the $SO(10)$ is broken into the SM gauge symmetry, the effective Yukawa interactions as obtained from Eq. (\ref{LY_GUT}) can be written as
\beqa \label{LY_THDM}
-{\cal L}_Y  & =& \overline{Q}_L \left( \left(\alpha_1 H + \alpha_2 F \right) h_d + \beta_1^* \tilde{H} \tilde{h}_u \right) d_R + \overline{Q}_L \left( \left(\beta_1 H + \beta_2 F \right) h_u + \alpha_1^* \tilde{H} \tilde{h}_d \right) u_R \nonumber \\
& + & \overline{L}_L \left( \left(\alpha_1 H - 3  \alpha_2 F \right) h_d + \beta_1^* \tilde{H} \tilde{h}_u \right) e_R + \overline{L}_L \left( \left(\beta_1 H - 3 \beta_2 F \right) h_u + \alpha_1^* \tilde{H} \tilde{h}_d \right) \nu_R\, \nonumber \\
& + & \frac{1}{2} \nu_R^T C^{-1} F \nu_R\, S\,+\, {\rm h.c.}\,, \eeqa
where $\tilde{h}_{u,d} = i \sigma_2 h_{u,d}^*$ and
\be \label{HF}
H = 2 \sqrt{2}\, Y_{10}\,,~~\tilde{H} = 2 \sqrt{2}\, \tilde{Y}_{10}\,,~~ F= - i 4 \sqrt{\frac{2}{3}}\, Y_{126}\,.\ee
It is seen from Eq. (\ref{LY_THDM}) that in the sub-GUT scale theory - three charged fermion Yukawa coupling matrices $Y_{u,d,e}$, the Dirac neutrino Yukawa coupling matrix $Y_\nu$ and the singlet neutrino mass matrix $M_R$ - all originate from only three fundamental Yukawa coupling matrices. In this way, the quark-lepton unification offered by $SO(10)$ GUTs is known to lead to a constrained and predictive framework for fermion masses and mixing parameters.

Additional assumptions can lead to even more predictive scenarios within the above framework \cite{Bajc:2005zf,Joshipura:2011nn}. An interesting and viable example is based on a Peccei-Quinn \cite{Peccei:1977hh} like $U(1)$ symmetry  under which $16_F \to e^{i \alpha}\, 16_F$ and $(10_H, \overline{126}_H) \to e^{-2 i \alpha}\,(10_H, \overline{126}_H)$. This leads to $\tilde{Y}_{10} = 0$ in Eq (\ref{LY_GUT}). A similar result can also be obtained if the model is embedded into a supersymmetric framework and the supersymmetry is broken at or above the GUT scale, see \cite{Buchmuller:2017vut} for example. The holomorphy of superpotential forbids Yukawa interactions with $10_H^*$ in this case. The effective Yukawa couplings in this class of frameworks can be parametrized using Eq. (\ref{LY_THDM}) as
\beqa \label{Eff_Yuk}
Y_d  &=& H^\prime + F^\prime \,, \nonumber \\
Y_u  &=& r\,(H^\prime + s F^\prime)\,, \nonumber \\
Y_e  &=& H^\prime - 3 F^\prime \,, \nonumber \\
Y_ \nu &=& r\,(H^\prime -3 s F^\prime)\,, \nonumber \\
M_R  &=& v_S^\prime\, F^\prime \,, \eeqa
where we have defined
\be \label{Eff_Yuk_def}
H^\prime= \alpha_1 H\,,~~F^\prime= \alpha_2 F\,,~~ r=\frac{\beta_1}{\alpha_1}\,,~~ s=\frac{\alpha_1 \beta_2}{\alpha_2 \beta_1}\,~~v_S^\prime = \frac{v_S}{\alpha_2}\,.\ee
Here, $v_S$ is VEV of the SM singlet field $S$. The SM fermion mass spectrum including that of RH neutrinos can be obtained from just two Yukawa coupling matrices in this case. Both $h_u$ and $h_d$ are required to be light through additional fine-tuning. The Yukawa sum rules obtained in Eq. (\ref{Eff_Yuk}) are identical to the one obtained in the minimal suspersymmetric $SO(10)$ model \cite{Babu:1992ia,Clark:1982ai,Aulakh:1982sw,Aulakh:2003kg} with an important difference that the effective theory below the GUT scale  is two-Higgs-doublet model (2HDM) of type II in the present case.

Alternatively, if only $h_d$ is assumed light instead of two Higgs doublets, the effective theory below the GUT scale is described by Eq. (\ref{LY_THDM}) with $\beta_{1,2} = 0$. The sub-GUT scale theory is the SM with $Y_{u}$, $Y_{d}$ and $Y_{e}$ as uncorrelated coupling matrices. The Dirac and Majorana neutrino couplings are obtained as $Y_\nu = Y_u$ and $M_R \propto Y_d - Y_e$. In comparison, this scenario is less predictive than the one discussed earlier in terms of the number of Yukawa couplings and we shall not consider here in subsequent discussion.

The complete unification of quarks and leptons of a given generation and choice of minimal Higgs sector allows one to determine the masses of RH neutrinos in the given framework. Assuming that the light neutrino masses are generated dominantly through type I seesaw mechanism\footnote{The $\overline{126}_H$ also contains a weak triplet sub-multiplet whose VEV can induce masses for the light neutrinos through type II seesaw mechanism. We assume that this contribution is sub-dominant compared to type I as it is known to lead to inconsistent neutrino mass spectrum in the minimal model considered here \cite{Joshipura:2011nn}.}, the SM neutrino mass matrix is given by
\be \label{Mnu}
M_\nu = - v_u^2\, Y_\nu\,M_R^{-1}\, Y_\nu^T\,, \ee
where $v_{u,d} = \langle h_{u,d} \rangle$, $v_u^2 + v_d^2 \equiv v^2 = (174\, {\rm GeV})^2$ and $v_u/v_d \equiv \tan\beta$. If no finely tuned cancellations are assumed between the $Y_{10}$ and $Y_{126}$ contributions in Eq. ({\ref{Eff_Yuk}), one finds
\be \label{Ynu_Yu}
(Y_\nu)_{ij} = c_{ij}\,(Y_u)_{ij}\,,\ee
where $c_{ij}$ are numerical factors of ${\cal O}(1)$. The RH neutrino masses can be approximated by inverting Eq. (\ref{Mnu}) and substituting Eq. (\ref{Ynu_Yu}). Simplification assuming only the third generation gives an approximate magnitude of the heaviest RH neutrino mass as 
\be \label{M3}
M_{N_3} \simeq 8 \times 10^{13}\,{\rm GeV}\, \times c_{33}^2 \times \left( \frac{\sin\beta}{0.83}\right)^2 \times \left(\frac{y_t}{0.44}\right)^2 \times \left(\frac{0.05\,{\rm eV}}{m_{\nu_3}}\right)\,. 
\ee
Here, $y_t$ is the value of top quark Yukawa coupling at the GUT scale. Further, Eq. (\ref{LY_THDM}) also gives 
\be \label{MR_hier}
M_R = \frac{v_S^\prime}{4}\, (Y_d - Y_e)\, ,\ee
which implies that the hierarchy among the masses of RH neutrinos is similar to that of the down-type quarks or charged leptons. 
\beqa \label{MN_hierarchy}
\frac{M_{N_2}}{M_{N_3}} & \simeq & {\cal O}\left( \frac{m_s}{m_b}, \frac{m_\mu}{m_\tau}\right) \simeq (2\, {\rm -} \, 6) \times 10^{-2}\,, \nonumber \\
\frac{M_{N_1}}{M_{N_2}} & \simeq & {\cal O}\left( \frac{m_d}{m_s}, \frac{m_e}{m_\mu}\right) \simeq (0.5\, {\rm -}\, 5) \times 10^{-2}\,.\eeqa

Typically, Eqs. (\ref{M3},\ref{MN_hierarchy}) imply the following approximate ranges for the masses of RH neutrinos in the model. 
\be \label{MN_range}
M_{N_1} \simeq 10^{9\,{\rm -}\, 11}\, {\rm GeV}\,,~~M_{N_2} \simeq 10^{10\,{\rm -}\, 12}\, {\rm GeV}\,,~~M_{N_3} \simeq 10^{12\,{\rm -}\, 14}\, {\rm GeV}\,.\ee
The spectrum of RH neutrinos is hierarchical although the inter generational hierarchy is not as strong as that in the up-type quark sector.

\section{Leptogenesis}
\label{sec:leptogenesis}
We now discuss the generation of lepton asymmetry through decays of heavy RH neutrinos within this model. Starting from the most general DME, we obtain a relatively simple analytical expression applicable for the kind of mass spectrum of RH neutrinos given by Eq. (\ref{MN_range}) and for the hierarchical Dirac neutrino Yukawa coupling matrix $Y_\nu$ given by Eq. (\ref{Ynu_Yu}).  

\subsection{Density Matrix Equations}
\label{sec:leptogenesis_A}
In thermal leptogenesis, the lepton asymmetry gets generated by CP violating out-of-equilibrium  decays of the RH neutrinos. CP asymmetry arises through the interference between the tree and 1-loop diagrams involving heavy neutrinos decaying into leptons and Higgs. For $M_{N_1} \gg 10^{12}$ GeV, the combination of lepton and anti-lepton flavor states which couple to $N_i$ can be treated as states which maintain the coherence between their production and inverse decays. For example, one can define these states as
\beqa \label{Li}
| L_i \rangle = \sum_\alpha C_{i \alpha}\, |L_\alpha \rangle\,,~~~ | \bar{L}_i \rangle = \sum_\alpha \bar{C}_{i \alpha}\, |\bar{L}_\alpha \rangle\,,
\eeqa
with $C_{i \alpha} = \langle L_i | L_\alpha \rangle$ and $\bar{C}_{i \alpha} = \langle \bar{L}_i | \bar{L}_\alpha \rangle$ and they can be explicitly determined from the vertex between $N_i$, $L_\alpha$ and Higgs. In general $C_{i \alpha} \neq \bar{C}_{i \alpha}$. However, if only tree level contribution is considered then
\be \label{C}
C_{i \alpha} = \bar{C}_{i \alpha} = \frac{Y_{i \alpha}}{\sqrt{(Y^\dagger Y)_{ii}}}\,. \ee
Here, $Y$ is the Yukawa coupling matrix $Y_\nu$ but in the basis of diagonal $Y_e$ and $M_R$. Explicitly, 
\be \label{Y}
Y = U_e^\dagger\, Y_\nu\, U_{\nu_R}\,,\ee
where $U_e$ and $U_{\nu_R}$ are obtained from the diagonalization of $Y_e$ and $M_R$, respectively. In this unflavored regime, the lepton asymmetry can be computed using the classical BE and final asymmetry is dominantly produced by the processes involving lightest RH neutrino \cite{Bodeker:2020ghk}. 

The coherent evolution of the states $|L_i\rangle$ and $|\bar{L}_i\rangle$ breaks down for $M_{N_1} < 10^{12}$ GeV because the charged leptons Yukawa interactions come into thermal equilibrium and the inverse decay processes start differentiating between the different flavors of leptons \cite{Vives:2005ra,DiBari:2005st,Abada:2006fw,Abada:2006ea,Nardi:2006fx,Engelhard:2006yg}. To account for these effects, the simple BE need to be replaced by the more general DME. These equations have been systematically derived in \cite{Blanchet:2011xq} and they can be used to evaluate asymmetry for the general mass spectrum of RH neutrinos. We list them below for the convenience of readers as well as for setting up the notations to be used in the subsequent discussion and analysis.
\beqa \label{dme}
\frac{d N_{N_j}}{dz} & = & -D_j  \left(N_{N_j} - N_{N_j}^{\rm eq}\right) \,, \nonumber \\
\frac{d N_{\alpha \beta}}{dz} & = & \sum_{j} \left[ \varepsilon_{\alpha \beta}^{(j)}\,D_j \left(N_{N_j} - N_{N_j}^{\rm eq}\right) -\frac{1}{2} W_j \left\{P^{(j)},N\right\}_{\alpha \beta} \right] \nonumber \\
& - & \frac{{\rm Im}(\Lambda_\tau)}{H\, z}\, \left(\delta_{\alpha 1} N_{1\beta} + \delta_{\beta 1} N_{\alpha 1} - 2 \delta_{\alpha 1} \delta_{\beta 1} N_{11}\right) \nonumber \\
& - & \frac{{\rm Im}(\Lambda_\mu)}{H\, z}\, \left(\delta_{\alpha 2} N_{2\beta} + \delta_{\beta 2} N_{\alpha 2} - 2 \delta_{\alpha 2} \delta_{\beta 2} N_{22}\right)\,,\eeqa
where $z = M_{N_1}/T$, $N_{N_j}$ ($N$) is a product of number density of $N_j^{\rm th}$ neutrino ($B-L$ asymmetry) and the comoving volume occupied by a heavy neutrino in ultra-relativistic thermal equilibrium. $N^{\rm eq}_{N_i}$ is the corresponding equilibrium value 
\be \label{Neq}
N_{N_i}^{\rm eq} = \frac{1}{2}\, x_i\, z^2\, {\cal K}_2(z_i)\,,\ee
such that $N_{N_i}^{\rm eq}(z_i \simeq 0) = 1$. Here, 
\be \label{xj}
x_j = \frac{M_j^2}{M_1^2}\,,~~z_j = \sqrt{x_j}\,z\,,\ee
and ${\cal K}_i(z)$ are modified Bessel functions of the second kind. $D_j$ is rescaled decay rate given by
\be \label{Dj}
D_j \equiv D_j(z) = K_j x_j z \frac{{\cal K}_1(z_j)}{{\cal K}_2(z_j)}\,,\ee
with
\be \label{Kjal}
K_{j \alpha} = \frac{M_j\, |Y_{\alpha j}|^2}{8 \pi\,H(M_j)}\,,~~K_j = \sum_\alpha K_{j \alpha} = \frac{M_j\, (Y^\dagger Y)_{jj}}{8 \pi\, H(M_j)}\,,\ee
and \be
\label{H}
H(z) = 1.66\, \sqrt{g_*}\,\frac{M_1^2}{M_P}\,\frac{1}{z^2}\,,\ee
is Hubble expansion rate.

The CP asymmetry in the decay of $j^{\rm th}$ RH neutrino in full three flavor regime is described by the matrix $\varepsilon_{\alpha \beta}^{(j)}$ in Eq. (\ref{dme})  and its explicit expression in terms of the Yukawa couplings and RH neutrino masses is given by \cite{Blanchet:2011xq,Beneke:2010dz}
\beqa \label{eps}
\varepsilon_{\alpha \beta}^{(j)} & = & \frac{3 i}{32 \pi \left(Y^\dagger Y \right)_{jj}}\,\sum_{i\neq j}\Big[ \frac{\xi (x_i/x_j)}{\sqrt{x_i/x_j}} \left(Y_{\alpha j} Y^*_{\beta i} (Y^\dagger Y)_{ij} - Y^*_{\beta j} Y_{\alpha i} (Y^\dagger Y)_{ji} \right) \nonumber \Big. \\
& + & \Big. \frac{2}{3(x_i/x_j -1)} \left( Y_{\alpha j} Y^*_{\beta i} (Y^\dagger Y)_{ji} - Y^*_{\beta j} Y_{\alpha i} (Y^\dagger Y)_{ij} \right) \Big]\,,\eeqa
where 
\be \label{xiz}
\xi(x) = \frac{2}{3} x\,\left[(1+x)\,\ln\left( \frac{1+x}{x}\right) - \frac{2-x}{1-x}\right]\,.\ee
$W_j$ represents an appropriately rescaled rate of the washout of $B-L$ symmetry and it is given by 
\be \label{Wj}
W_j \equiv W_j(z) =\frac{1}{4}K_j \sqrt{x_j}\,z_j^3\,{\cal K}_1(z_j) \,.\ee
Further,
\be \label{P}
P^{(j)}_{\alpha \beta} = C_{j \alpha} C^*_{j \beta} = \frac{Y_{\alpha j}\,Y^*_{\beta j}}{(Y^\dagger Y)_{jj}}\,, \ee
denotes the projection matrix describing how a particular combination of flavored asymmetry gets washed out. Finally, the last two terms in Eq. (\ref{dme}) account for damping in the off-diagonal terms of the $B-L$ asymmetry matrix $N$.  They are determined as
\beqa \label{ImLambda}
\frac{{\rm Im}(\Lambda_\mu)}{H\, z} & = & \frac{8 \times 10^{-3}\, y_\mu^2\, T}{H z} = 1.7 \times 10^{-10}\,\frac{M_P}{M_1}\,,\nonumber \\
\frac{{\rm Im}(\Lambda_\tau)}{H\, z} & = & \frac{8 \times 10^{-3}\, y_\tau^2\, T}{H z} = 4.7 \times 10^{-8}\,\frac{M_P}{M_1}\,.
\eeqa
When the temperature goes below $10^{12}$ GeV, the $y_\tau$ dependent interactions come into thermal equilibrium and leads to decoherence of $\tau$-lepton states. Subsequently, when the temperature drops below $10^9$ GeV the similar effects arise from the $y_\mu$ dependent interactions. Similarly, the electron Yukawa dependent damping term needs to be added if one considers $M_{N_1} < 10^6$ GeV.

The DME can be solved numerically to obtain the value of matrix $N$ at $z \gg 1$. The trace of the elements of the obtained matrix gives the final $B-L$ asymmetry
\be \label{NBL_def}
N^{\rm f}_{B-L} = \sum_\alpha N_{\alpha \alpha}\,.\ee 
Finally, the baryon to photon ratio can be determined using
\be \label{etaB}
\eta_B = 0.96 \times 10^{-2}\, N^{\rm f}_{B-L}\,,\ee
where the numerical factor accounts for $B-L$ asymmetry converted to baryon asymmetry through sphaleron interactions and dilution due to an increase in the number of photons in a comoving volume \cite{Laine:1999wv}.

\subsection{From DME to BE to an analytical solution}
\label{sec:leptogenesis_B}
The DME in Eq. (\ref{dme}) describe kinetic evolution of $B-L$ asymmetry for general spectrum of RH neutrino masses with $M_{N_1} > 10^6$ GeV. For the hierarchical Dirac neutrino Yukawa couplings and RH neutrino mass spectrum, such as Eqs. (\ref{Ynu_Yu},\ref{MN_range}) predicted in the present framework, it is possible to simplify the DME in terms of the Boltzmann equations to get an approximate analytical solution for $B-L$ asymmetry.

Since $Y_\nu$ is proportional to $Y_u$ in this model, the Dirac neutrino coupling matrix in the diagonal basis of the charged leptons and RH neutrinos is generically as hierarchical as the up-type quark mass matrix. The hierarchical structure of $Y$ implies
\be \label{eps_hrc}
\varepsilon_{\tau \tau}^{(j)} \gg \varepsilon_{\alpha \beta}^{(j)}\,,~~~\alpha\,{\rm or}\,\beta \neq \tau\,,\ee
from Eq. (\ref{eps}). Therefore, the $B-L$ asymmetry is largely determined by the $(3,3)$ element of the matrix $N$. Moreover, for hierarchical $Y_\nu$,
\be \label{P3}
P^{(3)}_{\tau \tau} = |C_{33}|^2 = \frac{|Y_{33}|^2}{(Y^\dagger Y)_{33}}\simeq 1\,, \ee
from Eq. (\ref{P}).

The first round of asymmetry gets generated from the production of $N_3$ and its subsequent decays in a narrow interval of  $T$ around $T_{B3} \sim M_{N_3}$. Since $M_{N_3} \gsim 10^{12}$ GeV in the present framework, the flavor effects can be ignored. With this, only $N_3$ dependent terms in Eq. (\ref{dme}) contribute. This along with Eqs. (\ref{eps_hrc},\ref{P3}) leads to the  following BE for asymmetry evolution: 
\beqa \label{N3}
\frac{d N_{\tau \tau}}{dz} & = & \varepsilon_{3 \tau}\,D_3 \left(N_{N_3} - N_{N_3}^{\rm eq}\right) - W_3\, N_{\tau \tau}\,,\eeqa
where
\be \label{vareps_jalpha}
\varepsilon_{j \alpha} \equiv \varepsilon_{\alpha \alpha}^{(j)}\,.\ee
The solution of the above equation can be expressed analytically as \cite{Buchmuller:2004nz}
\be \label{N3_sol}
N_{\tau \tau} (T \simeq T_{B3}) = \varepsilon_{3 \tau}\, \kappa(K_{3 \tau})\,\ee
where $ \kappa(K_{3 \tau})$ is efficiency factor. In the case of initial thermal abundance of RH neutrinos, it is given by 
\be \label{kappa}
\kappa(x) = \frac{2}{x z_B(x)} \left(1-\exp\left(-\frac{1}{2} x z_B(x)\right)\right)\,,
\ee
with
\be \label{zB}
z_B(x) = 2 + 4\, x^{0.13}\, \exp\left(-\frac{2.5}{x} \right)\,.\ee
Note that more explicit expression of $T_{B3}$ is given by $T_{B3} = M_{N_3}/z_B(K_3)$ \cite{Blanchet:2011xq}.  The asymmetry produced at this stage is to be used as the initial condition while solving for kinetic equations involving $N_2$ production and decay and evolution of corresponding $B-L$ asymmetry.

Next, $N_2$ production at $T \sim T_{B2} = M_2/z_B(K_2)$ can be captured by $N_2$ dependent terms in Eq. (\ref{dme}). Since $M_{N_2} < 10^{12}$ GeV the flavor dependent effects become important as asymmetry in $\tau$-lepton flavor evolves differently than the other ones. The matrix $N$ can be decomposed in two components: one for asymmetry in the flavor state $|L_\tau \rangle \equiv |\tau \rangle$  and the other in its orthogonal combination which we denote by $| \tau^\perp \rangle$. The relevant BE can be deduced from Eq. (\ref{dme}) as
\beqa \label{N2}
\frac{d N_{\tau \tau}}{dz} & = & \varepsilon_{2 \tau}\,D_2 \left(N_{N_2} - N_{N_2}^{\rm eq}\right) - P_{2 \tau}\, W_2\, N_{\tau \tau}\,,\nonumber \\
\frac{d N_{\tau^\perp \tau^\perp}}{dz} & = & \varepsilon_{2 \tau^\perp}\,D_2 \left(N_{N_2} - N_{N_2}^{\rm eq}\right) - P_{2 \tau^\perp}\, W_2\, N_{\tau^\perp \tau^\perp}\,,\eeqa
where $\varepsilon_{2 \tau^\perp} = \varepsilon_{2 \mu} + \varepsilon_{2 e}$, $P_{2 \tau^\perp} = P_{2 \mu} + P_{2 e}$ and $P_{j \alpha} \equiv P^{(j)}_{\alpha \alpha}$. The above two equations are decoupled and their solutions are given by
\beqa \label{N2_sol}
N_{\tau \tau}(T \simeq T_{B2})& = & \varepsilon_{2 \tau}\, \kappa(K_{2 \tau})\,+\, \varepsilon_{3 \tau}\, \kappa(K_{3 \tau})\, e^{-\frac{3 \pi}{8} K_{2 \tau}}\,, \nonumber \\
N_{\tau^\perp \tau^\perp}(T \simeq T_{B2})& = & \varepsilon_{2 \tau^\perp}\, \kappa(K_{2 \tau^\perp})\,,\eeqa
where $K_{2 \tau^\perp} = K_{2 \mu} + K_{2 e}$. The first terms in the above equations are analogous to analytical solution used earlier as Eq. (\ref{N3_sol}). The second term in $N_{\tau \tau}$ is an initial asymmetry produced at $T \simeq T_{B3}$ along with the exponential factors which quantify the washout by processes involving $N_2$ when temperature comes down to $T \simeq T_{B2}$.

Finally, at $T$ around $T_{B1} = M_1/z_B(K_1)$, the production of $N_1$ is considered. Since $M_{N_1} > 10^9$ GeV in the present model, $\mu$-lepton Yukawa interactions are still out-of-equilibrium at this temperature and one can proceed with two flavor case as before. The kinetic evolution is governed by equations similar to Eq. (\ref{N2}) with $N_2$ replaced by $N_1$. Explicitly,
\beqa \label{N1}
\frac{d N_{\tau \tau}}{dz} & = & \varepsilon_{1 \tau}\,D_1 \left(N_{N_1} - N_{N_1}^{\rm eq}\right) - P_{1 \tau}\, W_1\, N_{\tau \tau}\,,\nonumber \\
\frac{d N_{\tau^\perp \tau^\perp}}{dz} & = & \varepsilon_{1 \tau^\perp}\,D_1 \left(N_{N_1} - N_{N_1}^{\rm eq}\right) - P_{1 \tau^\perp}\, W_1\, N_{\tau^\perp \tau^\perp}\,,\eeqa
where $\varepsilon_{1 \tau^\perp} = \varepsilon_{1 \mu} + \varepsilon_{1 e}$ and $P_{1 \tau^\perp} = P_{1 \mu} + P_{1 e}$. The above equations can be solved taking into account the asymmetry produced from $N_2$ decays given by Eqs. (\ref{N2_sol}) as initial conditions. For this purpose, the $N_{\tau^\perp \tau^\perp}(T \simeq T_{B2})$ needs to be decomposed into a parallel and an orthogonal to vector $| \tau_1^\perp \rangle$. Here, the subscript ``$1$" is used to denote that $|\tau_1^\perp \rangle$ is normalized component, orthogonal to $| L_\tau \rangle$, of the state $|L_1 \rangle$ which couples to $N_1$.  It is different from $|\tau^\perp \rangle$ which is a component of the state $|L_2 \rangle$ which couples to $N_2$. Using Eq. (\ref{Li}), their explicit expressions are given by
\beqa \label{perp}
|\tau_1^\perp \rangle = \frac{1}{\sqrt{|C_{1e}|^2+|C_{1\mu}|^2}}\, \left(C_{1e}\, |L_e \rangle + C_{1 \mu}\, |L_\mu \rangle \right)\,, \nonumber \\
|\tau^\perp \rangle = \frac{1}{\sqrt{|C_{2e}|^2+|C_{2\mu}|^2}}\, \left(C_{2e}\, |L_e \rangle + C_{2\mu}\, |L_\mu \rangle \right)\,.\eeqa
The projection of $|\tau^\perp \rangle$ on the $|\tau_1^\perp \rangle$ is then obtained as 
\be \label{p12}
p_{12} \equiv |\langle \tau_1^\perp | \tau^\perp \rangle|^2 = \frac{|C_{2 \mu}^* C_{1 \mu} + C_{2 e}^* C_{1 e}|^2}{\left(|C_{2 \mu}|^2 + |C_{2 e}|^2 \right) \left(|C_{1 \mu}|^2 + |C_{1 e}|^2 \right)}
\ee

A component, parallel to $|\tau_1^\perp \rangle$, of the asymmetry $N_{\tau^\perp \tau^\perp}(T \simeq T_{B2})$ produced at the $N_2$ stage is washed out by subsequent processes involving $N_1$ as it can be seen from the second in Eq. (\ref{N1}). Therefore, the approximate analytical solution of Eq. (\ref{N1}) together with initial conditions provided by Eq. (\ref{N2_sol}) are given by
\beqa \label{N1_sol}
N_{\tau \tau}(T \simeq T_{B_1})& = & \varepsilon_{1 \tau} \kappa(K_{1 \tau}) + \varepsilon_{2 \tau} \kappa(K_{2 \tau})\, e^{-\frac{3 \pi}{8} K_{1 \tau}} + \varepsilon_{3 \tau} \kappa(K_{3 \tau})\, e^{-\frac{3 \pi}{8} (K_{2 \tau}+K_{1 \tau})}\,, \nonumber \\
N_{\tau^\perp \tau^\perp}(T \simeq T_{B_1})& = & \varepsilon_{1 \tau^\perp} \kappa(K_{1 \tau^\perp}) + p_{12}\, \varepsilon_{2 \tau^\perp} \kappa(K_{2 \tau^\perp})\,e^{-\frac{3 \pi}{8} K_{1 \tau^\perp}} + (1 - p_{12})\, \varepsilon_{2 \tau^\perp} \kappa(K_{2 \tau^\perp})\,. \eeqa
The first term in both the equations above is the usual solution of Eq. (\ref{N1}) while the subsequent terms are asymmetries produced from earlier stages with appropriate wash-out factors. The last term in the second equation is the $N_2$ produced asymmetry which escapes wash-out effects from $N_1$ interactions due to non-trivial flavor effects. At $T < T_{B1}$, $N_1$ production gets suppressed, washout interactions go out of equilibrium and the asymmetry given by Eq. (\ref{N1_sol}) gets frozen. The final $B-L$ asymmetry is then given by 
\be \label{N_BL}
N^{\rm f}_{B-L} \simeq  N_{\tau \tau}(T \simeq T_{B_1}) + N_{\tau_1^\perp \tau_1^\perp}(T \simeq T_{B_1})\,. \ee

Eqs. (\ref{N_BL}, \ref{N1_sol}) represent an approximate analytical solution of the DME for RH neutrino masses and neutrino Yukawa couplings as predicted in the model. Further simplification is possible to achieve by inspecting the magnitudes of various terms in Eq. (\ref{N1_sol}). From the seesaw relation between the light neutrino masses, $Y$ and $M_{N_i}$ and from Eq. (\ref{Kjal}), we find that typically
\beqa \label{K_ialpha}
K_{2 \tau} & = & \frac{M_P}{8 \pi \times 1.66\, \sqrt{g_*}}\frac{|Y_{32}|^2}{M_{N_2}} \simeq \frac{M_P}{8 \pi \times 1.66\, \sqrt{g_*}}\frac{m_{\nu_3}}{v_u^2} > 67.8\,, \nonumber \\
K_{1 \tau} & = & \frac{M_P}{8 \pi \times 1.66\, \sqrt{g_*}}\frac{|Y_{31}|^2}{M_{N_1}} > \frac{M_P}{8 \pi \times 1.66\, \sqrt{g_*}}\frac{m_{\nu_2}}{v_u^2} > 11.7\,, \nonumber \\
K_{2 \mu} & = & \frac{M_P}{8 \pi \times 1.66\, \sqrt{g_*}}\frac{|Y_{22}|^2}{M_{N_2}} \simeq \frac{M_P}{8 \pi \times 1.66\, \sqrt{g_*}}\frac{m_{\nu_2}}{v_u^2} > 11.7\,, \nonumber \\
K_{1 \mu} & = & \frac{M_P}{8 \pi \times 1.66\, \sqrt{g_*}}\frac{|Y_{21}|^2}{M_{N_1}} > \frac{M_P}{8 \pi \times 1.66\, \sqrt{g_*}}\frac{m_{\nu_1}}{v_u^2} > 2.7\,. \eeqa
The second equality in the above equations follows from $m_{\nu_i} \approx {\cal O} \left(v_u^2 |Y_{ij}|^2/M_{N_j}\right)$ which is obtained from the seesaw formula neglecting the neutrino mixing. We also make use of the hierarchical structure of $Y$, i.e. $|Y_{31}| > |Y_{21}|$ and $|Y_{21}| > |Y_{11}|$. Subsequently, $m_{\nu_3} \ge \sqrt{\Delta m_{\rm atm}^2}$, $m_{\nu_2} \ge \sqrt{\Delta m_{\rm sol}^2}$ and $m_{\nu_1} \ge 2\, {\rm meV}$ are used to derive the respective lower bounds. These values of neutrino masses are in accordance with the spectrum predicted by the model as will be shown in the next section. Consequently, all the terms in Eq. (\ref{N1_sol}) containing exponentials are generically negligible since all the relevant $K_{i \alpha} \gg 1$. Further, we find $\varepsilon_{1 \alpha} \ll \varepsilon_{2 \alpha}$ for hierarchical $Y_\nu$. The final asymmetry is then dominantly given by a rather simple formula
\be \label{N_BL_approx}
N^{\rm f}_{B-L} \simeq  (1 - p_{12})\, \varepsilon_{2 \tau_2^\perp}\, \kappa(K_{2 \tau_2^\perp})\,. \ee

It is seen the final $B-L$ asymmetry is a fraction of the one generated during $N_2$ production  which does not get erased by flavor dependent $N_1$ interactions due to $p_{12} \neq 1$. This clearly implies the $N_2$ dominated leptogenesis as already advocated by \cite{DiBari:2008mp,DiBari:2010ux,Buccella:2012kc,DiBari:2014eya,Fong:2014gea,DiBari:2015oca,DiBari:2017uka,DiBari:2020plh} in $SO(10)$-inspired models. The expression of final asymmetry, Eq. (\ref{N_BL_approx}), is very similar to the one derived earlier in \cite{Antusch:2011nz} for minimal type I seesaw model with two RH neutrinos. In the case of the latter, the total asymmetry turns out to be almost independent of $M_{N_2}$. The presence of $N_3$ in this $SO(10)$ framework does not have significant direct implication on the $B-L$ asymmetry due to the strong wash-out effects at the subsequent stages as discussed before. However, $N_3$ allows the $N_2$  CP asymmetry to remain proportional to $M_{N_2}$ unlike in \cite{Antusch:2011nz}. With already negligibly small $N_1$ induced contribution, this makes the final asymmetry more or less independent of $M_{N_1}$.

Note that various analytical solutions used to derive Eq. (\ref{N_BL_approx}) are known to match with the exact numerical solutions of the respective BE within a difference of a factor of ${\cal O}(1)$. The $N^{\rm f}_{B-L}$ in  Eq. (\ref{N_BL_approx}), therefore, does not give a very accurate number for the baryon asymmetry. Nevertheless, its extremely simple form allows us to fit the baryon to photon ratio together with the other observables of fermion masses and mixing parameters through $\chi^2$ minimization method in a computationally very efficient way. We also note that while deriving the above result, we did not take into account the effects of the so-called phantom terms \cite{Blanchet:2011xq,Antusch:2010ms} which could be of importance particularly in the flavored leptogenesis case. Nevertheless, these effects are included in full DME which we solve numerically for viable solutions as described in detail in the next section.

\section{Numerical Analysis and Results}
\label{sec:numerical}
We carry out a numerical investigation for the viability of the model in explaining fermion masses and mixing parameters along with the observed baryon to photon ratio. This is done using the $\chi^2$ optimization method also followed earlier in \cite{Joshipura:2011nn,Buchmuller:2017vut} for a similar analysis. The function is defined as
\be \label{chs}
\chi^2 = \sum_i \left( \frac{O^{\rm th}_i - O^{\rm exp}_i}{\sigma_i} \right)^2\,, \ee
where $O^{\rm th}_i$ is the  theoretically computed value of $i^{\rm th}$ observable in the model,  $O^{\rm exp}_i$ is the corresponding experimental value extrapolated at the GUT scale $M_{\rm GUT}$ and $\sigma_i$ is the uncertainty in $O^{\rm exp}_i$.  The list of observables includes a total of 19 quantities: 9 charged fermion masses (or diagonal Yukawa couplings), 2 neutrino squared mass differences, 4 parameters of quark mixing matrix, 3 mixing angles in the neutrino sector and a baryon to photon ratio, $\eta_B$. The last observable is computed using Eq. (\ref{etaB}) and an analytical expression of $N_{B-L}^{\rm f}$, Eq. (\ref{N_BL_approx}).

The values $O^{\rm exp}_i$ for the charged fermion Yukawa couplings and CKM parameters are obtained by evolving their low energy values to $M_{\rm GUT}$ using renormalization group equations. The detail of this procedure is given in Appendix \ref{App:RGE}. As the effective theory below $M_{\rm GUT}$ contains a pair of Higgs doublet, the extrapolation is carried out assuming 2HDM between $M_t$ and $M_{\rm GUT}$. We also choose $\tan\beta = 1.5$ which is a favorable value if the underlying non-supersymmetric sub-GUT scale theory arises from supersymmetric theory \cite{Mummidi:2018nph,Mummidi:2018myd, SuryanarayanaMummidi:2020ydm}. Small $\tan\beta$ is preferred by the stability of electroweak vacuum and Higgs mass constraint in this class of models. For the light neutrino masses, we assume normal ordering. It is known that for normal hierarchy in neutrino masses and for low $\tan\beta$ the renormalization group induced correction to neutrino masses and leptonic mixing parameters are small. We, therefore, use the low energy values for these observables from \cite{Esteban:2020cvm} as $O^{\rm exp}_i$. For baryon to photon ratio $\eta_B$, we use the value from \cite{Planck:2018nkj}. Values of various $O^{\rm exp}_i$ obtained in the aforementioned ways are listed in the third column of Table \ref{tab:bestfit}. For $\sigma_i$, we follow the same convention used in \cite{Buchmuller:2017vut} and consider a $30 \%$ standard deviation in the values of light quark Yukawa couplings ($y_u$, $y_d$ and $y_s$) and $\eta_B$ and $10 \%$ standard deviation in all the remaining observables. These conservative standard deviations are taken in order to account for next-to-leading order RGE effects and threshold corrections.

$O^{\rm th}_i$ for Yukawa couplings and mixing parameters are computed from Eqs. (\ref{Eff_Yuk},\ref{Mnu}) following the usual diagonalization procedure and the standard parametrization of mixing matrices as given in \cite{ParticleDataGroup:2020ssz}. For $\eta_B$, we use Eq. (\ref{etaB}) along with the analytical result obtained in Eq. (\ref{N_BL_approx}). All these quantities are non-linear functions of a small set of parameters. One can perform an overall rotation on three flavors of $16_F$ to make $H^\prime$ in Eq. (\ref{Eff_Yuk}) real and diagonal. $F^\prime$ remains symmetric in the new basis and both these matrices can be parametrized in terms of 15 real parameters. Further, $r$ and $v_S^\prime$ can be chosen real without loss of generality while $s$ is complex in general. Altogether, these 19 parameters are used to reproduce the 19 observables as discussed earlier using the $\chi^2$ function minimization. Note that even if the number of parameters and observables are the same here, it is not guaranteed that all the observables can be reproduced as the latter are complex non-linear functions of the original parameters and there exist several correlations between the observables.

At the minimum of $\chi^2$, we evaluate predicted values of various observables which are not yet measured. This includes the Dirac and Majorana CP phases in the lepton sector, the mass of the lightest neutrino $m_{\nu_1}$, the effective mass of ordinary $\beta$-decay, $m_\beta = \sqrt{\sum_i |U_{ei}|^2 m_{\nu_i}^2}$, and neutrinoless double $\beta$-decay, $m_{\beta \beta} = |\sum_i U_{ei}^2\, m_{\nu_i}|$, and the mass spectrum of the RH neutrinos. For the definition of CP phases, we use the PDG convention \cite{ParticleDataGroup:2020ssz}. There already exists indirect constraint on the values of leptonic Dirac phase from global fits of neutrino oscillation data \cite{Esteban:2020cvm}. However, we do not include it in the $\chi^2$ function as the allowed range at $3\sigma$ is still considerably wide. We rather compute this phase at the minimum of $\chi^2$ for a large number of points to derive its comprehensive prediction in the considered $SO(10)$ model.    

\subsection{Best fit solution}
We obtain the best fit solution corresponding to $\chi^2 = 1.7$ at the minimum. The corresponding results and predictions are listed in Table \ref{tab:bestfit}.
\begin{table}[!ht]
	\begin{center} 
		\begin{math} 
			\begin{tabular}{cccc}
				\hline
				\hline 
				~~~Observable~~~  & ~~~~~~~~~$O_i^{\rm th}$~~~~~~~~~  &~~~ ~~~~~~$O_i^{\rm exp}$~~~~~~~~~  & $P_i = (O_i^{\rm th} - O_i^{\rm exp})/\sigma_i$ \\
				\hline
				$y_u$  & $2.91 \times 10^{-6}$ & $2.91\times 10^{-6}$ & 0.0 \\
				$y_c$   & $1.49\times10^{-3}$& $1.47\times 10^{-3}$& 0.1\\
				$y_t$   & $0.437 $ & $0. 443$& -0.1\\
				$y_d$  & $ 3.46 \times 10^{-6} $ & $5.04\times10^{-6}$  & -1.0 \\
				$y_s$   & $0.87 \times10^{-4}$ & $1.01\times10^{-4}$ & -0.4 \\
				$y_b$   & $5.30 \times10^{-3}$ & $5.40 \times10^{-3}$& -0.2 \\
				$y_e$   & $2.18 \times 10^{-6}$ & $2.16 \times 10^{-6}$ & 0.1 \\
				$y_{\mu}$  &$4.68 \times10^{-4}$ & $4.51\times10^{-4}$ & 0.4\\
				$y_{\tau}$   & $7.75 \times10^{-3}$ & $7.63 \times10^{-3}$& 0.2\\
				$\Delta m^2_{\text{sol}}\, [{\rm eV}^2]$ & $7.48\times10^{-5}$& $7.42\times10^{-5}$& 0.1\\
				$\Delta m^2_{\text{atm}}\, [{\rm eV}^2]$ & $2.517\times10^{-3}$& $2.517\times10^{-3}$& 0.0\\
				$|V_{us}|$ & 0.2352 & 0.2321 & 0.1\\
				$|V_{cb}|$ & 0.0393 & 0.0399 & -0.2 \\
				$|V_{ub}|$ & 0.0036 & 0.0036 & 0.0 \\
				$\sin\delta_{\rm CKM}$ & 0.924 & 0.931 & -0.1  \\
				$\sin^2 \theta _{12}$ $(\theta_{12})$ & 0.311 (33.90$^{\circ}$) & 0.304 (33.44$^{\circ}$) & 0.2\\
				$\sin^2 \theta _{23}$ $(\theta_{23})$ & 0.554 (48.1$^{\circ}$)  & 0.573 (49.2$^{\circ}$) & -0.3 \\
				$\sin^2 \theta _{13}$ $(\theta_{13})$ & 0.02229 (8.59$^{\circ}$)  & 0.02219 (8.57$^{\circ}$) & 0.0 \\		
				$\eta_B$& $6.10 \times 10^{-10}$ & $6.12\times 10^{-10}$ & -0.1\\		  	
				\hline
				\multicolumn{4}{c}{Predictions} \\
				\hline
				$\delta_{\rm PMNS}$[$^{\circ}$] & 354.6 & $M_{N_1}$ [GeV]  & $4.36 \times 10^{9}$ \\
				$\alpha _{21}$ [$^{\circ}$] & 181.8  & $M_{N_2}$ [GeV]  & $1.97 \times 10^{11}$ \\
				$\alpha _{31}$ [$^{\circ}$] & 123.7 & $M_{N_3}$ [GeV]  & $ 8.86 \times 10^{11}$ \\
				$m_{\nu _1}$ [eV] &0.0060  & \\
				$ m_{\beta }$  [eV]  &0.0108  & & \\	
				$m_{\beta \beta}$  [eV]  & 0.0082 &\\						
				\hline
				\hline 
			\end{tabular}
		\end{math}
	\end{center}
	\caption{Results and predictions obtained for the best fit solution corresponding to $\chi^2 = 1.7 $ at the minimum.} 
	\label{tab:bestfit} 
\end{table}
All the 19 observables are fitted within $1 \sigma$ of the desired ranges as it can be seen from the Table. The largest deviation is found in $y_d$ which deviates from the GUT scale extrapolated experimental value by $30 \%$. This is in agreement with the results of older fit \cite{Joshipura:2011nn} in which deviation in $y_d$ is also found to be largest. The best fit solution predicts RH neutrino masses more or less in the ranges already anticipated from the simplified analytical arguments in section \ref{sec:YukawaSector}.

As mentioned earlier the value of $\eta_B$ in Table \ref{tab:bestfit} is computed using the analytical expression of $B-L$ asymmetry, Eq. (\ref{N_BL_approx}). Since we now have all the relevant Yukawa couplings and masses of the RH neutrinos determined for the best fit, we can solve the DME numerically to obtain the exact value of $B-L$ asymmetry. The solutions for the number densities of RH neutrinos and diagonal and off-diagonal components of $B-L$ asymmetry are given as a function of $z$ in Fig. \ref{fig1}. 
\begin{figure}[t]
\centering
\subfigure{\includegraphics[width=0.49\textwidth]{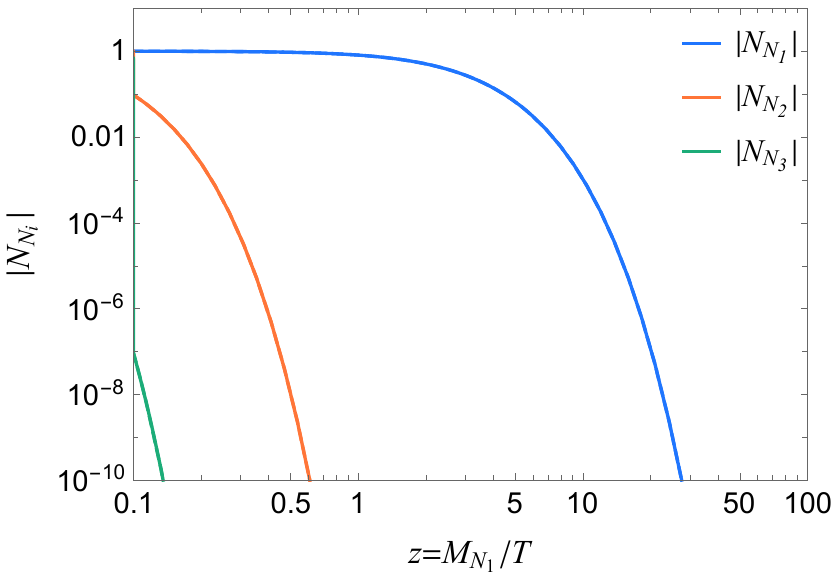}}
\subfigure{\includegraphics[width=0.49\textwidth]{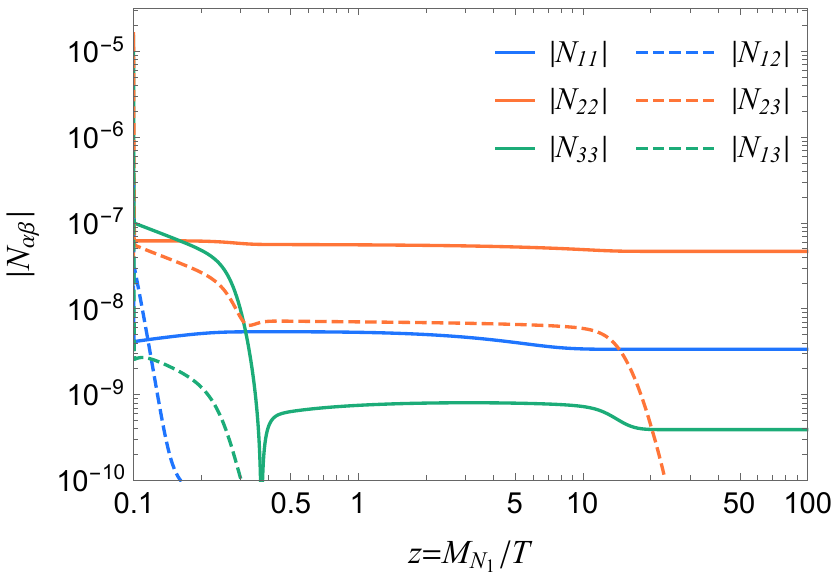}}
\caption{Solutions obtained by solving density matrix equations, Eq. (\ref{dme}), for number densities of the RH neutrinos (left panel) and flavored $B-L$ asymmetry (right panel) for the best fit point given in Table \ref{tab:bestfit}.} 
\label{fig1}
\end{figure}
We have assumed thermal initial abundance for RH neutrino number densities\footnote{Since $K_2  \gg 1$ (see, Eqs. (\ref{Kjal},\ref{K_ialpha})), the final asymmetry will be the same even if vanishing initial abundance for the RH neutrinos is assumed.}. For the off-diagonal components, the temperature evolution of the $|N_{\alpha \beta}|$ is identical to that of $|N_{\beta \alpha}|$ as the CP asymmetry matrices $\varepsilon^{(j)}$ in Eq. (\ref{dme}) are hermitian. All the off-diagonal components of $B-L$ asymmetry get damped rapidly as it can be seen from the right panel in Fig. \ref{fig1}. This happens due to the hierarchical structure of $Y$ as discussed earlier in the section \ref{sec:leptogenesis_B}.

It can be seen from Fig. \ref{fig1} that the $N_{22}$ component of asymmetry dominates over all the others at $z \gg 1$. This component is dominantly generated from decays of $N_2$ and remains more or less constant as the temperature decreases. Only a small part of this asymmetry gets washed out by the subsequent processes involving $N_1$ if no specific flavor alignment is assumed. The asymmetry generated by $N_3$ decay goes through the strong wash-out while that induced by $N_1$ decays remains small. Therefore, the final asymmetry is dominantly given by number density $|N_{22}|$ which favors the $N_2$ dominated scenario. The value of baryon to photon ratio evaluated from this exact numerical solution is obtained as
\be \label{etaB_num}
\eta^{\rm num}_B = 0.96 \times 10^{-2}\, \sum_\alpha N_\alpha = 4.2 \times 10^{-10}\,.
\ee
The above value differs from the analytical value given in Table \ref{tab:bestfit} by $30\%$. This is expected as various analytical approximations taken to derive simplified expression of $N_{B-L}^{\rm f}$ in Eq. (\ref{N_BL_approx}) are known to give rise to a difference of ${\cal O}(1)$ factor.  Nevertheless, given $10 \%$ to $30 \%$ deviations in various observables used to determine the Yukawa couplings and RH neutrino masses we also allow $30 \%$ standard deviation in $\eta_B$ from its experimentally measured value $\eta_B^{\rm exp} = (6.12 \pm 0.04) \times 10^{-10}$  \cite{Planck:2018nkj} and the obtained numerical value in Eq. (\ref{etaB_num}) is still well within this conservative range.

\subsection{Predictions for leptonic observables}
To derive the detailed predictions of the model for the various observables in the lepton sector, we go beyond the best fit solution and carry out a comprehensive search for other minima of $\chi^2$. Since the $\chi^2$ function includes $n=19$ observables, we consider all the solutions with $\chi^2/n \le 1$ at their local minimum as acceptable solutions. Once these solutions are obtained, we compute the corresponding $\eta_B^{\rm num}$ by numerically solving the full DME and define 
\be \label{delta_etaB}
\delta \eta_B \equiv \left| \frac{\eta_B - \eta_B^{\rm num}}{\eta_B^{\rm num}}\right|\,.
\ee
Recall that $\eta_B$ is evaluated using the simple analytical expression, Eq. (\ref{N_BL_approx}), of $N_{B-L}^{\rm f}$ and it is already included in the $\chi^2$ function. We then consider solutions with $\chi^2/n \le 1$ and $\delta \eta_B \le 1$ as valid solutions and compute various relevant observables. The results are displayed in Figs. \ref{fig2} and  \ref{fig3}. 
\begin{figure}[t]
\centering
\subfigure{\includegraphics[width=0.48\textwidth]{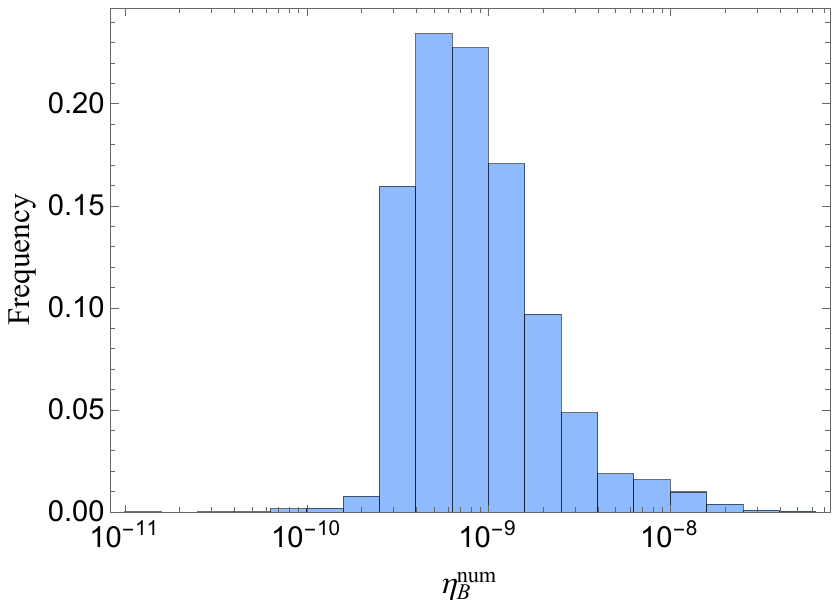}}
\subfigure{\includegraphics[width=0.49\textwidth]{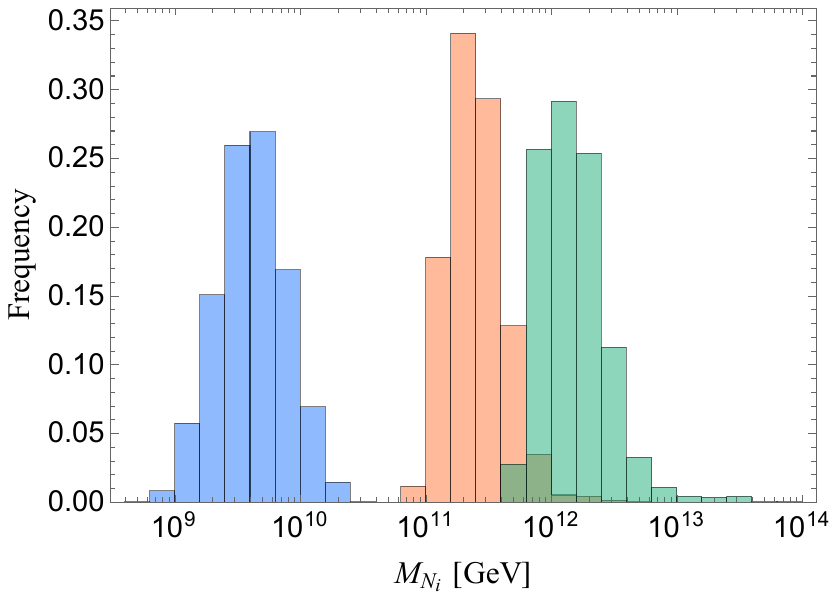}}
\caption{Probability distributions for $\eta_B^{\rm num}$ (left panel) and the masses of RH neutrinos (right panel) obtained for the solutions with $\chi^2/n \le 1$ and $\delta \eta_B \le 1$.} 
\label{fig2}
\end{figure}
\begin{figure}[!ht]
\centering
\subfigure{\includegraphics[width=0.48\textwidth]{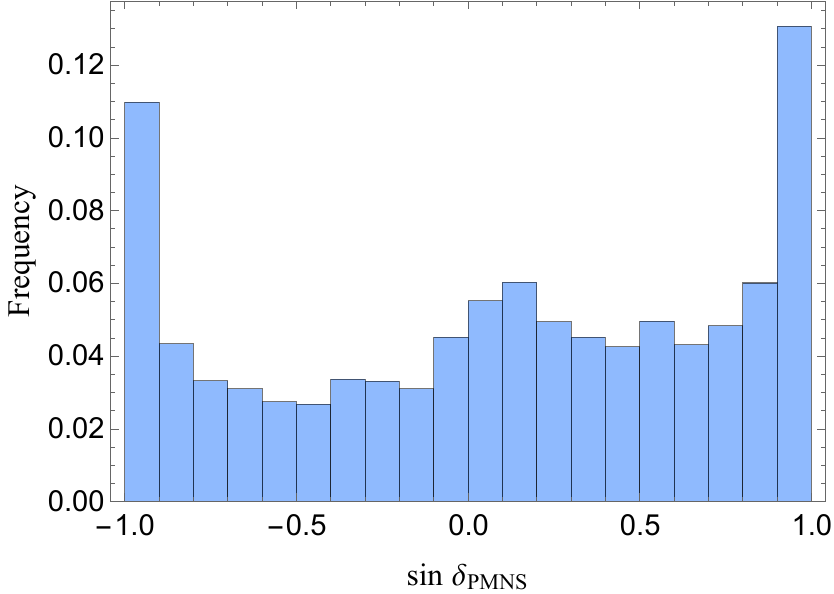}}
\subfigure{\includegraphics[width=0.48\textwidth]{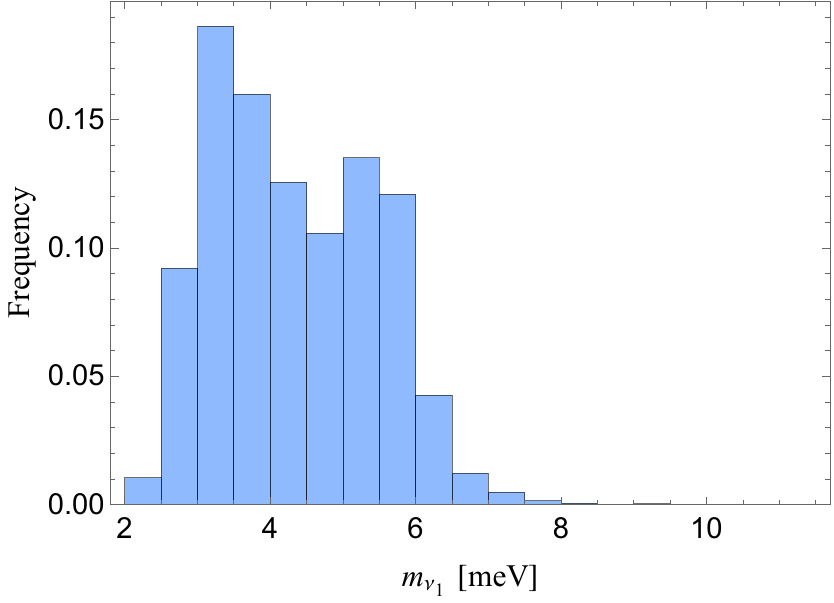}}
\subfigure{\includegraphics[width=0.48\textwidth]{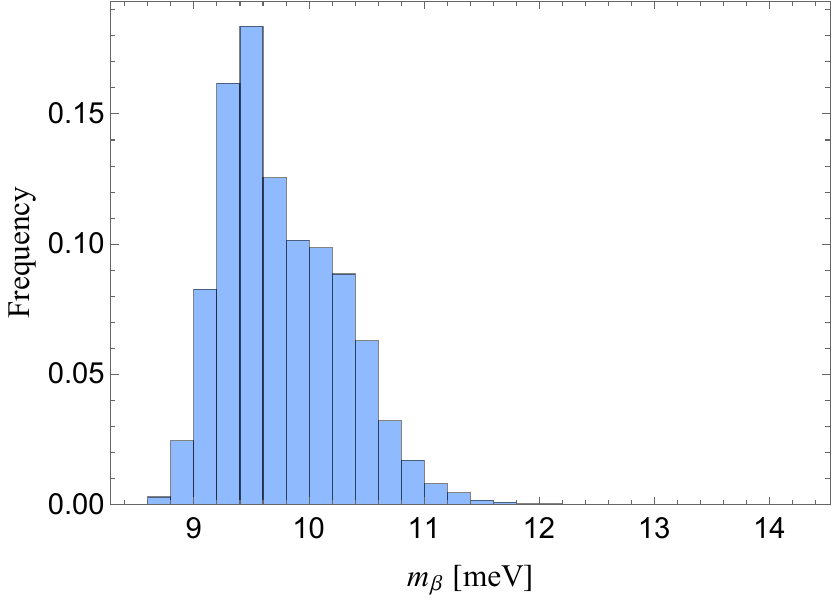}}
\subfigure{\includegraphics[width=0.48\textwidth]{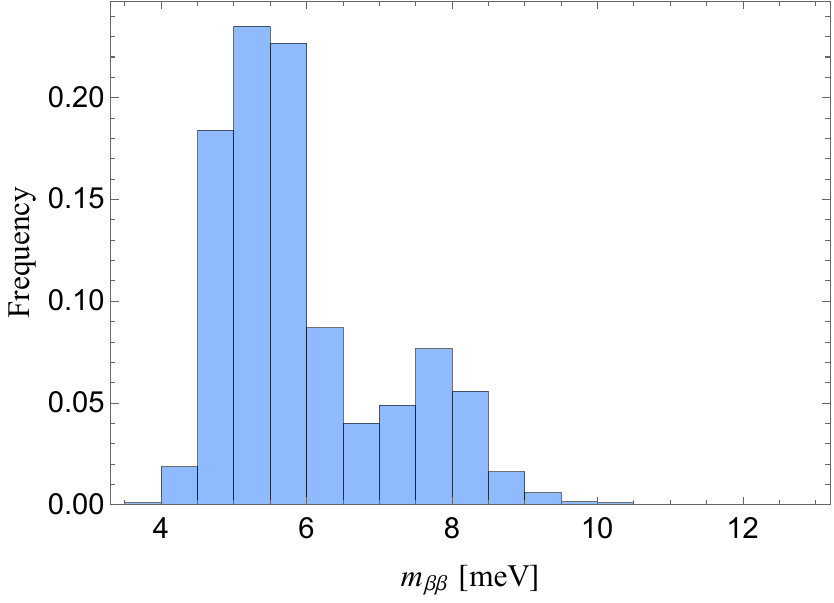}}
\caption{Probability distributions for the leptonic observables: $\sin \delta_{\rm PMNS}$ (top-left), mass of the lightest neutrino (top-right), the effective mass of beta decay (bottom-left), and of neutrinoless double beta decay (bottom-right) obtained from solutions with $\chi^2/n \le 1$ and $\delta \eta_B \le 1$.} 
\label{fig3}
\end{figure}

The numerically computed value of the baryon to photon ratio is found between $10^{-10}$ and $10^{-8}$ for more than $98 \%$ of the total valid points. The masses of the RH neutrinos are predicted in a narrow range as seen from the right panel in Fig. \ref{fig2} and they are in agreement with the typical values given in Eq. (\ref{MN_range}) which are predicted from the $SO(10)$ Yukawa relations. The inter-generational hierarchy between their masses is of the similar order of the charged lepton or down type quark masses. The predictions for the leptonic sector observables which can be verified in the ongoing and future experiments are given in Fig. \ref{fig3}. The model does not prefer any particular value of the leptonic Dirac CP phase and almost all the values of $\delta_{\rm PMNS}$ between $0$ and $2 \pi$ are equally favoured. Interestingly, viable leptogenesis can be achieved even with vanishing $\delta_{\rm PMNS}$ in which case the Majorana CP phases and CP violation in the RH neutrino sector can reproduce the observed baryon asymmetry. The mass of the lightest neutrino lies between $2$-$8$ meV, $m_\beta$ is predicted between $9$-$11$ meV while the effective mass of neutrinoless double beta decay is found between $4$-$9$ meV. Moreover, we also find very specific correlations between the Dirac CP phase and one of the Majorana phases as well as between $m_{\nu_1}$ and $m_{\beta \beta}$. These are displayed in Fig. \ref{fig4}. 
\begin{figure}[t]
\centering
\subfigure{\includegraphics[width=0.47\textwidth]{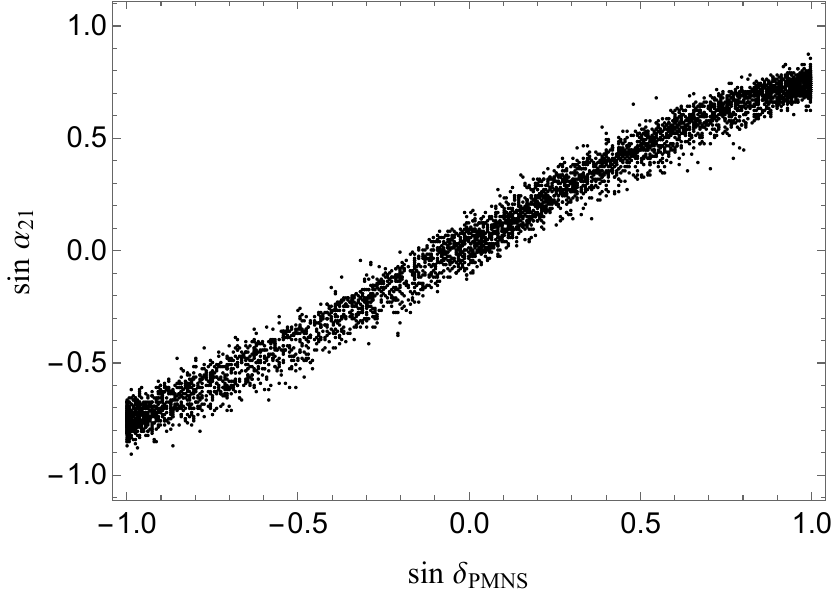}}
\subfigure{\includegraphics[width=0.49\textwidth]{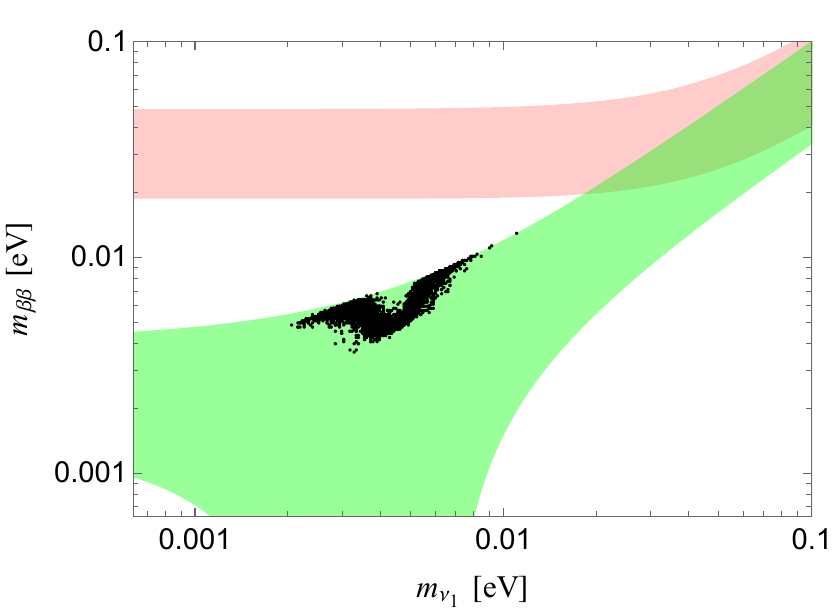}}
\caption{The black points show correlations between $\delta_{\rm PMNS}$ and $\alpha_{21}$ (left panel) and between $m_{\nu_1}$ and $m_{\beta \beta}$ (right panel) as predicted in the model. In the right panel, the green and red bands indicate the generic regions allowed by normal and inverted ordering of the light neutrino masses, respectively.}
\label{fig4}
\end{figure}
The noteworthy feature of the underlying model is that it predicts values of $m_{\nu_1}$ and $m_{\beta \beta}$ which are not vanishingly small. All these predictions and correlations make the model falsifiable.

\section{Summary and Discussion}
\label{sec:summary}
The $SO(10)$ GUTs offer an ideal platform to study leptogenesis as the RH neutrinos, whose out-of-equilibrium decays generate the lepton asymmetry, are unified along with the other standard model matter fields in this class of theories. In the concrete $SO(10)$ models with minimal scalars in the Yukawa sector, it is possible to use this quark-lepton unification to determine completely or partially the RH neutrino mass spectrum as well as the Dirac neutrino Yukawa couplings from the observed fermion mass spectrum. We have carried out this investigation in an $SO(10)$ model with complex $10_H$ and $\overline{126}_H$ in the Yukawa sector. Additional symmetry is used to determine all the quark and lepton masses in terms of only two Yukawa coupling matrices. Using these correlations, we first estimate the expected range of RH neutrino masses. Starting from the most general DME that describe the kinetic evolution of number densities of RH neutrinos and flavor specific lepton asymmetries, a simplified analytical solution is approximated for the obtained ranges of RH neutrino masses. This expression of asymmetry is then used to fit the observed value of the baryon to photon ratio together with all the fermion masses and mixing parameters in the underlying model.

Several solutions with statistically acceptable fits are obtained and their predictions for leptonic observables, such as the Dirac and Majorana CP phases, absolute neutrino mass scale, the effective mass for ordinary beta and neutrinoless double beta decay amplitudes and RH neutrino masses, are derived. For all these solutions, we also numerically solve the full DME to ensure that the analytical solution for lepton asymmetry used in the fit is not significantly different from the exact solution. It is seen that the analytical solution is in agreement with the latter within ${\cal O}(1)$ factor. It is seen that the successful leptogenesis in this model does not prefer any particular value of Dirac or Majorana CP phases. However, a specific correlation has been predicted between one of the Majorana and Dirac CP phases. The mass of the lightest neutrino is predicted to be in the range $2$-$8$ meV and $m_{\beta \beta}$ in $4$-$10$ meV. Some of these predictions evaluated in this concrete $SO(10)$ model differ significantly from the same obtained considering the generic $SO(10)$-inspired leptogenesis scenario \cite{DiBari:2020plh}. The differences in the results indicate the need for analysis of specific $SO(10)$ frameworks for the  precise determination of corresponding predictions.

For the analysis of fermion mass spectrum and leptogenesis, our focus has been on the Yukawa sector of non-supersymmetric $SO(10)$ frameworks as it is the most relevant. The complete model also requires analysis of the GUT symmetry breaking and gauge coupling unification. These aspects have been explored in \cite{Bertolini:2009qj,Bertolini:2009es,Bertolini:2012im}. The precision unification of gauge couplings consistent with the proton decay limit in non-supersymmetric frameworks requires presence of intermediate scale symmetry \cite{Bertolini:2009qj,Chakrabortty:2017mgi,Ernst:2018bib,Ohlsson:2020rjc} or large GUT scale threshold corrections \cite{Schwichtenberg:2018cka} or light matter fields which are incomplete representations of $SO(10)$ \cite{Patel:2011eh,Bhattacherjee:2017cxh}. These effects can modify the running of Yukawa couplings directly or indirectly through modification in the running of the gauge couplings. If the correction to the GUT scale Yukawa couplings due to these effects are smaller than $10$-$30\%$ standard deviations considered in the fits then the quantitative results and predictions derived in this work are expected to remain unchanged.    

\section*{Acknowledgements}
The work of KMP is partially supported by a research grant under INSPIRE Faculty Award (DST/INSPIRE/04/2015/000508) from the Department of Science and Technology, Government of India. The computational work reported in this paper was performed on the High Performance Computing (HPC) resources (Vikram-100 HPC cluster) at the Physical Research Laboratory, Ahmedabad.

\appendix
\section{Renormalization group evolution of Yukawa couplings}
\label{App:RGE}
In this Appendix, we discuss our method of extrapolation of Yukawa couplings from the weak to the GUT scale. We first obtain values of the gauge couplings and fermion masses at the top quark pole mass $M_t=173.1$ GeV by a procedure described in \cite{Mummidi:2018nph}. For the convenience of the readers, these values are provided here in Table \ref{tab:inputs}. 
\begin{table}[t]
	\begin{center}
		\begin{tabular}{|cc|cc|cc|cc|}
		\hline
		Parameter & Value & Parameter & Value & Parameter & Value & Parameter & Value\\
		\hline
		$g_1$ & 0.4632 & $m_u$ & 1.21 MeV & $m_d$ & 2.58 MeV & $m_e$ & 0.499 MeV \\
		$g_2$ & 0.6540 & $m_c$ & 0.61 GeV & $m_s$ &  52.74  MeV & $m_{\mu}$&0.104 GeV\\
		$g_3$ & 1.1630 & $m_t$ & 163.35 GeV & $m_b$ & 2.72 GeV & $m_{\tau}$&1.759 GeV \\
		\hline
		\end{tabular}
		\caption{Obtained values of the gauge couplings and fermion masses at renormalization scale $M_t = 173.1$ GeV in ${\overline{\rm MS}}$ scheme. See Appendix C of \cite{Mummidi:2018nph} for details.}
		\label{tab:inputs}
	\end{center}
\end{table}
From the masses of the charged fermions, respective Yukawa couplings are extracted as given by
\beqa \label{}
Y_u (M_t) &=& \frac{1}{v_u}\, \left( \ba{ccc}
    m_{u} & 0 & 0 \\
    0 & m_c & 0\\
    0 & 0 & m_t \ea \right)\,,\nonumber \\
Y_d (M_t) & = & \frac{1}{v_d} V_{\rm CKM} \left( \ba{ccc}
    m_{d} & 0  &  0\\
    0 & m_s & 0\\
    0 & 0 & m_b \ea \right)\,,\nonumber \\
Y_e(M_t) & = & \frac{1}{v_d} \left( \ba{ccc}
	m_{e} & 0 & 0 \\
	0 & m_\mu & 0 \\
	0 & 0 & m_\tau \ea \right)\,, \eeqa
where $v_d = v \cos\beta$, $v_u=v \sin\beta$ with $v=174 $ GeV  and $\tan\beta = 1.5$. This value of $\tan\beta$ is preferred in the models in which the sub-GUT scale non-supersymmetric two-Higgs doublet model arise from supersymmetric theory \cite{Mummidi:2018nph,Mummidi:2018myd}. $V_{\rm CKM}$ is the Cabibbo-Kobayashi-Maskawa (CKM) matrix given in terms of three mixing angles and a CP phase in the standard parametrization. We use the following values for these parameters
\be \label{CKM_prms} 
\sin\theta_{12} = 0.2265,~\sin\theta_{23} = 0.0405,~\sin\theta_{13} = 0.0036,~\delta_{\rm CKM} = 1.196, \ee 
at low scale and they are obtained from \cite{ParticleDataGroup:2020ssz}.

The above gauge and Yukawa couplings are evolved from $M_t$ to the GUT scale, $M_{\text{GUT}}=2\times 10^{16}$ GeV, using one-loop renormalization group equations (RGE) of type-II THDM which are given in Appendix A of \cite{Mummidi:2018nph} and reproduced below for the convenience of readers. The $\beta$-functions for gauge and Yukawa couplings are defined as
\be \label{diff_beta}
\mu\,\frac{d C}{d\mu} = \frac{1}{16 \pi^2}\,\beta_{C}^{(1)}\,, \ee
where $C$ stands for couplings and $\mu$ is the renormalization scale. Explicit expressions of $\beta$-functions are
\beqa \label{beta_g}
\beta_{g_1}^{(1)}  =  \frac{21}{5}\, g_{1}^{3}\,,~~~
 \beta_{g_2}^{(1)}  =   - 3\, g_{2}^{3}\,, ~~~
\beta_{g_3}^{(1)}  =  -7\, g_{3}^{3}\,,
\eeqa
and
\beqa \label{beta_Y}
\beta_{Y_u}^{(1)} & = & Y_u\, \Big(-8\, g_{3}^{2}  -\frac{17}{20}\, g_{1}^{2}  -\frac{9}{4}\, g_{2}^{2} + 3\, \mbox{Tr}\Big({Y_{u}^{\dagger}  Y_{u}}\Big)  \Big) + \frac{1}{2} \,\Big(3\,{Y_u  Y_{u}^{\dagger}  Y_u}  + {Y_d  Y_{d}^{\dagger}  Y_u  }\Big)\,, \nonumber  \\ 
\beta_{Y_d}^{(1)} & = &  Y_d\,\big(-8\, g_{3}^{2}  -\frac{1}{4}\, g_{1}^{2}  -\frac{9}{4}\, g_{2}^{2} +3\, \mbox{Tr}\Big({Y_d  Y_{d}^{\dagger}}\Big)  + \mbox{Tr}\Big({Y_e  Y_{e}^{\dagger}}\Big)\Big) \nonumber \\
& + & \frac{1}{2} \,\Big(3\,{Y_d  Y_{d}^{\dagger}  Y_d}  + {Y_u  Y_{u}^{\dagger}  Y_d  }\Big)\,, \nonumber \\
\beta_{Y_e}^{(1)} & =  & Y_e\, \Big(3\, \mbox{Tr}\Big({Y_d  Y_{d}^{\dagger}}\Big)  + \mbox{Tr}\Big({Y_e  Y_{e}^{\dagger}}\Big)  -\frac{9}{4} \Big(g_{1}^{2} + g_{2}^{2}\Big)\Big) + \frac{3}{2}\, Y_e  Y_{e}^\dagger  Y_e\,.
\eeqa

At $M_{\rm GUT}$, the diagonal Yukawa couplings corresponding to charged fermions are  obtained by diagonalizing the complex Yukawa matrices such that $Y_f^{\rm diag.} =U_f^\dagger Y_f V_f$ for $f=u,d,e$. The CKM matrix is obtained as $V_{\rm CKM} = U_u^\dagger U_d$. The extrapolated values of diagonal Yukawa couplings and CKM parameters are given in the third column in Table \ref{tab:bestfit} and used as $O_i^{\rm exp}$ in $\chi^2$ function, Eq. (\ref{chs}).

\section{Parameters corresponding to the best fit solution}
\label{App:Input_bestfit}
In this Appendix, we give the values of 19 real parameters appearing in Eq. (\ref{Eff_Yuk}) obtained for the best fit solution. They are:
\beqa \label{input_prm}
H^\prime &=& \left(
\begin{array}{ccc}
 0.00023 & 0 & 0 \\
 0 & -0.04811 & 0 \\
 0 & 0 & -5.79504 \\
\end{array}
\right) \times 10^{-3}\,, \nonumber \\
F^\prime &=&  \left(
\begin{array}{ccc}
 -0.0088+0.0178 i & 0.0475\, -0.0889 i & 0.4635\, +0.6797 i \\
 0.0475\, -0.0889 i & 1.1279\, +0.5108 i & -1.2218-2.5921 i \\
 0.4635\, +0.6797 i & -1.2218-2.5921 i & 5.4683\, -5.9856 i \\
\end{array}
\right)\times 10^{-4}\,, \nonumber \\
r & = & 77.4189\,,~~s = 0.3140 - 0.0282\, i\,~~v_S^\prime = 9.84 \times 10^{14}\,{\rm GeV}\,.\eeqa

Typically, one finds from Eq. (\ref{Eff_Yuk}):
\be \label{}
\frac{y_t}{y_b} \simeq r\, \tan\beta\,.\ee
Therefore, relatively large value of $r$ is required to fit $y_t/y_b$ ratio for small $\tan\beta$. The above values of $r$ and $s$ indicate that the light Higgs doublet $h_u$ ($h_d$) dominantly comes from $10_H$ ($\overline{126}_H$). 

The Yukawa coupling matrices $Y_f$ ($f=d,u,e,\nu$) evaluated using Eq. (\ref{Eff_Yuk}) and the above values of parameters are obtained as
\beqa \label{Ys_bestfit}
Y_d &=&  \left(
\begin{array}{ccc}
 -0.00065+0.00178 i & 0.00475\, -0.00889 i & 0.04635\, +0.06797 i \\
 0.00475\, -0.00889 i & 0.06468\, +0.05108 i & -0.12218-0.25921 i \\
 0.04635\, +0.06797 i & -0.12218-0.25921 i & -5.2482-0.59856 i \\
\end{array}
\right)\times 10^{-3}\,, \nonumber \\
Y_u &=& \left(
\begin{array}{ccc}
 0.00045 i & 0.00096\, -0.00227 i & 0.01275\, +0.01551 i \\
 0.00096\, -0.00227 i & -0.00872+0.00996 i & -0.03535-0.06034 i \\
 0.01275\, +0.01551 i & -0.03535-0.06034 i & -4.3666-0.15741 i \\
\end{array}
\right)\times 10^{-1}\,, \nonumber \\
Y_e &=&  \left(
\begin{array}{ccc}
 0.00286\, -0.00534 i & -0.01426+0.02668 i & -0.13906-0.20392 i \\
 -0.01426+0.02668 i & -0.38648-0.15325 i & 0.36655\, +0.77764 i \\
 -0.13906-0.20392 i & 0.36655\, +0.77764 i & -7.43553+1.79568 i \\
\end{array}
\right)\times 10^{-3}\,,  \nonumber \\
Y_\nu &=&  \left(
\begin{array}{ccc}
 0.0007\, -0.00135 i & -0.00288+0.0068 i & -0.03825-0.04653 i \\
 -0.00288+0.0068 i & -0.12283-0.02987 i & 0.10606\, +0.18102 i \\
 -0.03825-0.04653 i & 0.10606\, +0.18102 i & -4.84601+0.47224 i \\
\end{array}
\right)\times 10^{-1}\,. \eeqa
The resulting $Y_\nu$ and $Y_u$ are in agreement with the expectation asserted in Eq. (\ref{Ynu_Yu}). The eigenvalues of $Y_\nu$, obtained by following the same biunitary diagonalization procedure used for $Y_{u,d,e}$, are:
\be \label{YN_ev}
y_{\nu_1} = 8.11\times 10^{-5}\,,~~y_{\nu_2} = 1.30 \times 10^{-2}\,,~~y_{\nu_3} = 0.486\,.
\ee

\bibliography{references}
\end{document}